\documentclass[aps,pre,twocolumn,superscriptaddress,showpacs,showkeys,floatfix]{revtex4}

\usepackage{natbib}
\usepackage{graphicx}
\usepackage{subfigure}
\usepackage{multirow}

\begin{document}
\title{Spectral analysis of boundary layers in Rayleigh-B\'{e}nard convection}
 
\date{\today}
\author{Jos \surname{Verdoold}}
\email{j.verdoold@tudelft.nl}
\affiliation{Department of Multi-Scale Physics, Delft University of Technology, Lorentzweg 1, 2628 CJ Delft, The Netherlands}

\author{Maarten \surname{van Reeuwijk}}
\affiliation{Department of Civil and Environmental Engineering, Imperial College London, London SW7 2AZ, United Kingdom}

\author{Mark J. \surname{Tummers}}
\affiliation{Department of Multi-Scale Physics, Delft University of Technology, Lorentzweg 1, 2628 CJ Delft, The Netherlands}

\author{Harm J. J. \surname{Jonker}}
\affiliation{Department of Multi-Scale Physics, Delft University of Technology, Lorentzweg 1, 2628 CJ Delft, The Netherlands}

\author{Kemo \surname{Hanjali\'{c}}}
\affiliation{Department of Multi-Scale Physics, Delft University of Technology, Lorentzweg 1, 2628 CJ Delft, The Netherlands}
\affiliation{Department of Mechanics and Aeronautics, University of Rome, ``La Sapienza'', Rome, Italy}

\begin{abstract}
A combined experimental and numerical study of the boundary layer in a 4:1
aspect-ratio Rayleigh-B\'{e}nard cell over a four-decade range of Rayleigh
numbers has been undertaken aimed at gaining a better insight into the character
of the boundary layers. The experiments involved the simultaneous Laser Doppler
Anemometry (LDA) measurements of fluid velocity at two locations, i.e. in the
boundary layer and far away from it in the bulk, for Rayleigh numbers varying
between $1.6 \times 10^7$ and $2.4 \times 10^9$. In parallel, direct numerical
simulations (DNS) have been performed for the same configuration for Rayleigh
numbers between $7.0 \times 10^4$ and $7.7 \times 10^7$.
The temperature and velocity probability density functions and the power spectra
of the horizontal velocity fluctuations measured in the boundary layer and in
the bulk flow are found to be practically identical. Except for the smallest
Rayleigh numbers, the spectra in the boundary layer and in the bulk central
region are continuous and have a wide range of active scales. This indicates
that both the bulk and the boundary layers are turbulent in the $\textrm{Ra}$ number
range considered. However, molecular effects can still be observed and the
boundary layer does not behave like a classical shear-driven turbulent boundary
layer.
\end{abstract}



\maketitle

\section{Introduction}
Rayleigh-B\'{e}nard (RB) convection in a fluid trapped between two horizontal
plates of unequal temperature with the bottom wall being warmer than the top
wall, has long served as a paradigm of thermal convection. Despite numerous
studies (see e.g. overviews by \cite{Siggia1994,Kadanoff2001}), it continues to
pose challenges because of a number of still unresolved and controversial
issues. We recall that RB convection is characterised by the Rayleigh number
$\textrm{Ra}=\alpha g \Delta \Theta H^3 / ( \nu \kappa )$, and the Prandtl number $\textrm{Pr}=\nu
/ \kappa$, where $\alpha$ is the thermal expansion coefficient, $\Delta \Theta$
is the temperature difference between the cold and the hot plate, $H$ is the
vertical distance between the plates, $\nu$ is the kinematic viscosity, and
$\kappa$ is the thermal diffusivity. In practical RB set-ups the lateral size,
$W$, of the plates is finite. This is taken into account by a third
dimensionless number known as the aspect ratio $\Gamma=W/H$.

A topic of intense debate is the scaling law for the dimensionless heat
transfer, $\textrm{Nu} \sim \textrm{Ra}^\gamma $, especially for the range $\textrm{Ra} > 10^7$
\citep{Niemela2000}. Here $\textrm{Nu}$ is the Nusselt number $\textrm{Nu}=hH/\lambda=-H/(\Delta
\Theta) \partial \overline{\Theta}/\partial z|_\mathrm{wall}$, $h$ stands for
the convective heat transfer coefficient, $\lambda$ is the thermal conductivity,
and $\overline{\Theta}$ is the averagedtemperature. 
\cite{Heslot1987,Castaing1989,Wu1990} proposed scaling laws for the
heat transfer and temperature statistics in RB convection, which differed from
the classical theories, such as that of \cite{Howard1963}. In the search for
general scaling behaviour, \cite{Grossmann2000} presented a unifying theory
dividing the $\textrm{Ra},\textrm{Pr}$ plane into several regions, each with different scaling.
Crucial assumptions in this theory are the existence of a large-scale
circulation, also known as the wind, and a Blasius type laminar boundary layer.
The assumption that the boundary layers of Rayleigh-B\'{e}nard convection behave
as a laminar Blasius boundary layer is supported by several observations: (i)
the scaling of the friction factor $C_f$ is compatible with that of a laminar
boundary layer past a flat plate at moderate $\textrm{Ra}$ \citep{Chavanne1997,
Chavanne2001, Amati2005}; (ii) the typical Reynolds number $\textrm{Re}$ of the flow is
low at moderate $\textrm{Ra}$: for water $\textrm{Re} \approx 200$ at $\textrm{Ra}=10^7$ and $\Gamma=4$,
which is generally accepted to be too low to sustain turbulence.
\cite{Grossmann2000} argued that at large enough $\textrm{Ra}$ this laminar boundary
layer breaks down and becomes turbulent, presuming that the transition to a
turbulent boundary layer is shear triggered. The critical $\textrm{Ra}$ number for this
transition was estimated to be in the order of $10^{14}$ for $\textrm{Pr}=1$.

Despite this apparently convincing evidence, it is not clear how a boundary
layer which is highly unsteady due to continuous plume impingement and
detachment can behave quantitatively as a laminar boundary layer. This would
indicate that the plumes only cause unsteady disturbances and a consequent
time-dependence in the boundary layer but are passive otherwise.

The aim of this paper is to enhance the understanding of the dynamics of the
boundary layers by comparing velocity statistics in the boundary layer with
those determined in the bulk. It has widely been regarded that the main
prerequisite for unsteady velocity fluctuations to be qualified as turbulence,
in contrast to other uncorrelated, chaotic and random, fluctuations, is a power
spectrum that is continuous over a wide range of scales. For this reason we
focused on the analysis of spectral density functions (sdf), complemented with
the probability density functions (pdf), aimed at clarifying the nature of the
fluctuations in the boundary layer. For this purpose we use in parallel
complementary numerical and experimental techniques: direct numerical
simulations (DNS) are performed for the lower $\textrm{Ra}$ values, Laser Doppler
Anemometry (LDA) measurements are conducted for the medium and higher $\textrm{Ra}$
values. All experiments and simulations combined cover a $\textrm{Ra}$ range between
$7.0\times 10^4$ and $2.4\times 10^9$. In the midrange both techniques overlap
over about one decade.

\section{Methodology}
\subsection{Experiments}
\begin{figure}
\centering
\includegraphics[width=86mm]{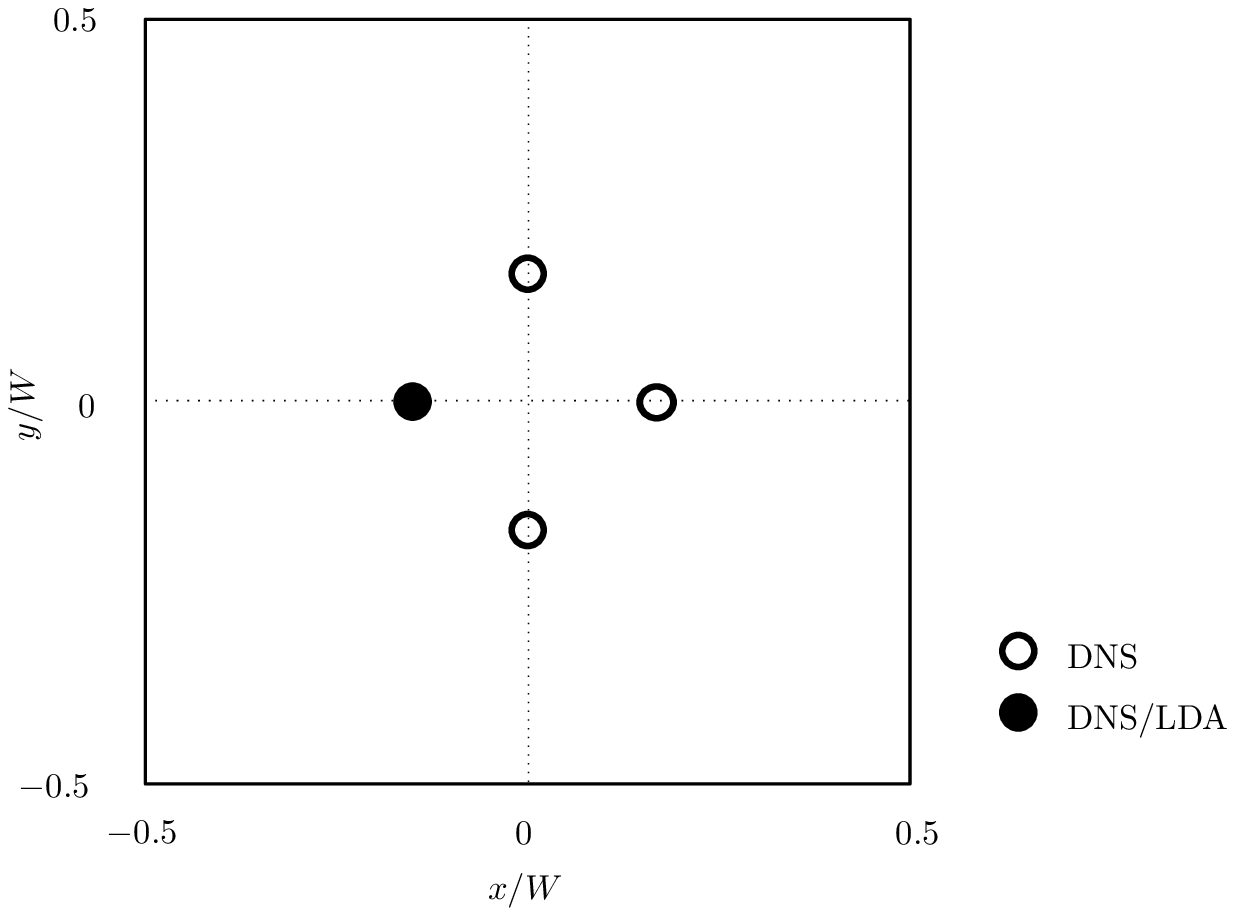}
\caption{\label{fig:measpoints}Schematic of the measurement points
in the RB cell for both DNS and LDA measurements}
\end{figure}

The experiments were conducted in a $600\times600\times155\ \textrm{mm}^3$
($W\times W\times H$) cell filled with water (described in more detail in
\cite{Verdoold2006}). In brief, two copper plates at the top and bottom are kept
at a constant temperature by passing water through the plates' internal
channels, drawn from two basins containing constant temperature water
(inaccuracy less than $0.03$~K). The plates thus impose a controlled temperature
difference $\Delta \Theta$ on the working fluid. Throughout this paper we will
use a Cartesian coordinate system with its origin at the centre of the bottom
wall, Fig.~\ref{fig:measpoints}. The $x$- and $y$- coordinates are measured
along the bottom wall, parallel to the side walls of the RB cell. The
$z$-coordinate measures the wall normal distance.

Velocities were measured by using two one-component laser Doppler anemometers
manufactured by Dantec. The green ($514.5$~nm) and blue ($488.0$~nm) colours of
a $5$~W Argon-ion laser are used to simultaneously measure the horizontal
component in the $x$-direction in the bulk and in the boundary layer. The
measured velocity component is in the direction of the large-scale circulation.
To enable the detection of instantaneous flow reversals one beam of each colour
was frequency pre-shifted by a Bragg cell. The light scattered by $5~\mu$m
polyamid seed particles was collected in on-axis forward direction, which, in
comparison to back-scatter, results in a superior signal-to-noise ratio of the
Doppler signals. The photomultiplier output signals were first downmixed
electronically to an effective pre-shift of 30~kHz and then fed to two Dantec
BSA signals processors to determine the instantaneous velocities. The LDA system
measured the instantaneous velocities of the seed particles with an inaccuracy
of $0.5$\%. The length and diameter of both measurement volumes are $0.8$~mm and
$0.1$~mm, respectively, which is sufficiently small to resolve the smallest
length scales of the flow. For each $\textrm{Ra}$ number, the measurement time $T$ was at
least $20$ hours to have sufficient data for an accurate statistical description
of the flow.

The mean data rate varied between $10$~Hz and $30$~Hz, depending on $\textrm{Ra}$ and
distance from the wall. For moderate $\textrm{Ra}$ numbers, the relatively low
characteristic velocities result in a relatively low data rate. Measurement
locations close to the wall have an inherently low data rate due to the low
fluid volume flux.

\renewcommand\arraystretch{1.7}
\renewcommand\tabcolsep{2.4mm}
\begin{table*}
\centering
\begin{tabular}{cr@{.}lr@{.}lr@{.}lccc}
$\textrm{Ra}$ & \multicolumn{2}{c}{$\textrm{Pr}$} &\multicolumn{2}{c}{$\Delta \Theta$ [K]} & \multicolumn{2}{c}{$\overline{\Theta}$ [$^\circ\mathrm{C}$]} & $\alpha\,\, [\textrm{K}^{-1}]$ & $ \nu\,\, [\textrm{m}^2\,\textrm{s}^{-1}]$ & T [s] \\
\hline
$1.6\times 10^7$    &   9&4 & 0&92  & 10&2  & $8.8\times 10^{-5}$ & $1.3\times 10^{-6}$ & $~1.1\times 10^5$ \\
$8.2\times 10^7$    &   7&0 & 1&57  & 19&9  & $2.1\times 10^{-4}$ & $1.0\times 10^{-6}$ & $~0.9\times 10^5$ \\
$4.2\times 10^8$    &   5&5 & 6&52  & 30&0  & $3.0\times 10^{-4}$ & $8.0\times 10^{-7}$ & $~1.2\times 10^5$ \\
$2.4\times 10^9$    &   5&5 & 25&1  & 30&3  & $3.0\times 10^{-4}$ & $8.0\times 10^{-7}$ & $~1.3\times 10^5$ \\
\hline
\end{tabular}
\caption{\label{tab:expdetails}Experimental details for the LDA measurements. 
All experiments have been performed in a $\Gamma=4$ aspect ratio domain. $T$
denotes the measurement time.}
\end{table*}
\renewcommand\arraystretch{1}

Table~\ref{tab:expdetails} shows the experimental parameters as well as the
relevant non-dimensional numbers. The Rayleigh number is varied by changing the
temperature difference $\Delta \Theta$ between the bottom and top plates from
nearly $1$~K at $\textrm{Ra}=1.6\times 10^7$ to $25$~K at $\textrm{Ra}=2.4 \times 10^9$. For the
highest $\textrm{Ra}$ numbers, the relatively large temperature differences cause larger
density differences, but these remain small enough to neglect non-Boussinesq
effects. The Prandtl number will change as well, but the ratio between the
smallest and largest $\textrm{Pr}$ number is only 1.7.

\begin{figure}
    \centering
    \includegraphics[width=\columnwidth]{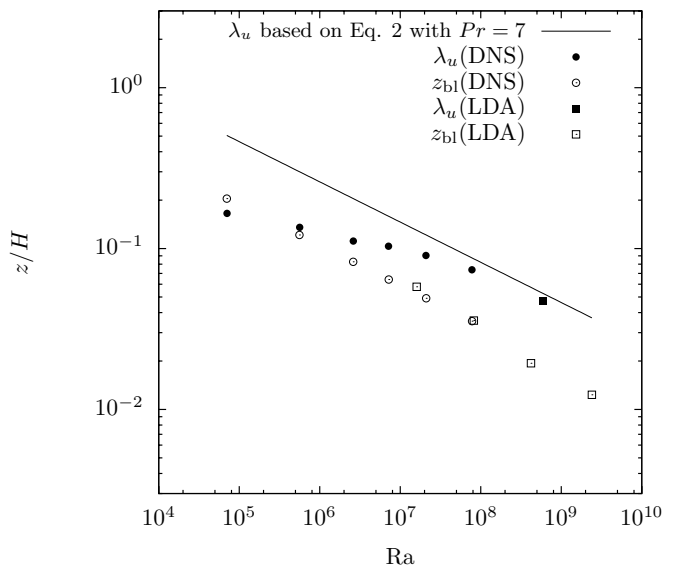}
    \caption{Kinematic boundary-layer height $\lambda_u$, based on
Equation~\ref{eq:Re_GL} as well as from DNS, and the measurement locations,
$z_\mathrm{bl}$,
    as function of $\textrm{Ra}$.}
    \label{fig:boundarylayerheights}
\end{figure}

\subsection{Simulations}
The direct numerical simulations cover the lower to moderate $\textrm{Ra}$ number range
between $\textrm{Ra}=7.0\times 10^4$ and $\textrm{Ra}=7.7 \times 10^7$. The numerics are based on
staggered finite differences, with a central second order scheme for advection,
and a second order Adams-Bashforth time integration scheme. To effectively
resolve the boundary layers, grid stretching in the wall-normal direction has
been implemented, and special care has been taken to preserve the skew-symmetry
of the advective operator \citep{Verstappen2003}. The code, as reported in
\cite{vanReeuwijk2007a}, has been extended to support sidewalls. In the
situation without sidewalls, the Poisson equation is Fourier-transformed in the
two homogeneous directions. As a result, the Poisson equation decouples for
these directions, and the pressure amplitudes can be solved per component by
solving a tridiagonal system in the wall-normal direction. To incorporate
sidewalls, the Fourier-transform was replaced by a cosine transform, which
automatically satisfies a Neumann boundary condition for pressure.
The sidewall boundary conditions for the velocity and temperature could be
enforced directly due to the explicit time-integration.

Dirichlet boundary conditions are applied at the top and bottom plate, i.e.\
no-slip and isothermal boundary conditions for velocity and temperature,
respectively. On the sidewalls, Neumann boundary conditions are applied for both
temperature and velocity, i.e.\ free-slip and insulating walls for velocity and
temperature. The free-slip (stress free) velocity boundary conditions were used
to avoid having to resolve the boundary layer at the sidewalls. Although this
changes the flow physics near the sidewalls, its global effect is limited for
large aspect ratio domains, as preliminary simulations showed. Neumann boundary
conditions were applied for pressure at all boundaries.

In the simulations, the properties of water are used except that $\textrm{Pr}=7$ is kept
fixed for all $\textrm{Ra}$.
The grid resolution (Table \ref{tab:simdetails}) is sufficient to resolve the
smallest turbulent scales, i.e. the Kolmogorov scale $\eta_{\mathrm{K}} =
(\nu^3/\varepsilon)^{1/4}$ and the Corrsin scale $\eta_\mathrm{\mathrm{C}} =
\textrm{Pr}^{-1/2}\ \eta_\mathrm{K}$. 
At $\textrm{Pr}=7$, the Corrsin scale dominates the grid requirements, and the resolution
roughly corresponds to $\Delta x \approx \Delta y \approx \Delta z \approx 4
\eta_\mathrm{C}$. 
To resolve the boundary layers at the top and bottom wall properly, grid
stretching has been applied in the wall-normal direction, such that at least
eight points are within the thermal boundary layer. 
It should be noted that the simulation at the highest $\textrm{Ra}$ number ($=7.7 \times
10^7$) may be slightly underresolved in the horizontal directions. 
High quality simulation as reported in \cite{Hanjalic2005} at $\textrm{Ra}=10^8$ and
$\textrm{Pr}=7$ required $768$ points per horizontal direction, but reduction to $384$
points was the only way to simulate over a sufficiently long time interval to
obtain reliable one-point statistics required for the comparison with the
experiments. 
The additional information shown in Table \ref{tab:simdetails} are the
dimensionless timestep $t/t^*$, the number of simulated turnovers $T/t^*$ and
the average Nusselt number $\textrm{Nu}$.
The typical convective turnover time $t^* \equiv H/U$ with $U$ the typical
velocity based on the free fall scaling.
The average Nusselt number is calculated as $\textrm{Nu} = - H / ( \Delta \Theta )
\left. \partial \overline{\Theta} / \partial z \right|_{\mathrm{wall}}$, with
$\overline{\Theta}$ averaged over the entire bottom surface and over many
turnovers.

\renewcommand\arraystretch{1.7}
\renewcommand\tabcolsep{4mm}
\begin{table}
\centering
\begin{tabular}{cccrr@{.}l}
$\textrm{Ra}$ & grid & $\Delta t / t^*$ & $T / t^*$ & \multicolumn{2}{c}{$\textrm{Nu}$} \\ 
\hline
$7.0 \times 10^4$ & $64^2 \times 48$   & $4.0 \times 10^{-4}$ & 240    & 3&8 \\ 
$5.6 \times 10^5$ & $96^2 \times 64$   & $2.3 \times 10^{-4}$ & 140    & 7&2 \\ 
$2.6 \times 10^6$ & $128^2 \times 92$  & $1.7 \times 10^{-4}$ & 165    & 11&3 \\ 
$7.2 \times 10^6$ & $192^2 \times 128$ & $1.3 \times 10^{-4}$ & 155    & 15&2 \\ 
$2.1 \times 10^7$ & $384^2 \times 160$ & $1.0 \times 10^{-4}$ & 40     & 20&0 \\ 
$7.7 \times 10^7$ & $384^2 \times 160$ & $1.0 \times 10^{-4}$ & 30     & 30&0 \\ 
\hline
\end{tabular}
\caption{ \label{tab:simdetails} Simulation details for DNS.}
\end{table}
\renewcommand\arraystretch{1}

\subsection{Measurement locations}
Simultaneous measurements have been performed in the boundary layer and in the
bulk of the cell for various $\textrm{Ra}$. The measurement location in the bulk is at
$x_{\mathrm{bulk}}=W/3$, $y_{\mathrm{bulk}}=W/2$ with a fixed wall distance
$z_{\mathrm{bulk}}=H/2$, see Fig.~\ref{fig:measpoints}. The measurement location
in the boundary layer has the same $x$- and $y$-coordinates, but the wall
distance is different, i.e., $z_{\mathrm{bl}}=0.4 \lambda_u$, where $\lambda_u$
is the kinematic boundary layer thickness determined as in \cite{Grossmann2000}: \begin{equation}
\lambda_u=H/\sqrt{\textrm{Re}}.
\label{eq:blheight}
\end{equation}
In this expression for $\lambda_u$ the Reynolds number, $\textrm{Re}$, was taken from
\cite{Grossmann2000,Grossmann2001}:
\begin{equation}
\textrm{Re}=0.073 \textrm{Ra}^{1/2}\textrm{Pr}^{-5/6}.
\label{eq:Re_GL}
\end{equation}

\begin{table*}
\centering
\begin{tabular}{ccr@{.}lcr@{.}lr@{.}l}
&$\textrm{Ra}$       &   \multicolumn{2}{c}{$\textrm{Pr}$} & $\textrm{\textrm{Re}}$ & \multicolumn{2}{c}{$\lambda_u/H$} & \multicolumn{2}{c}{$z_\mathrm{bl}/H$} \\
\hline
\multirow{6}{*}{\rotatebox{90}{DNS}}
&$7.0\times 10^4$    &   7&0 &   $3.8$           &   0&512                &
 0&204 \\
&$5.6\times 10^5$    &   7&0 &   $11$            &   0&305                &
 0&121 \\
&$2.6\times 10^6$    &   7&0 &   $23$            &   0&207                &  
0&0827 \\
&$7.2\times 10^6$    &   7&0 &   $39$            &   0&161                &  
0&0641 \\
&$2.1\times 10^7$    &   7&0 &   $66$            &   0&123                &  
0&0490 \\
&$7.7\times 10^7$    &   7&0 &   $1.3\times 10^2$ &   0&0884                &  
0&0354 \\
\hline
\multirow{4}{*}{\rotatebox{90}{LDA}}
&$1.6\times 10^7$    &   9&4 &   $48$            &   0&145                &
 0&0578 \\
&$8.2\times 10^7$    &   7&0 &   $1.2\times 10^2$ &   0&0897                &  
0&0357 \\
&$4.2\times 10^8$    &   5&5 &   $4.2\times 10^2$ &   0&0485                &
 0&0193 \\
&$2.4\times 10^9$    &   5&5 &   $1.0\times 10^3$ &   0&0309                &  
0&0123 \\
\hline
\end{tabular}
\caption{\label{tab:measlocations}Measurement heights for the
different $\textrm{Ra}$ experiments. The $\textrm{Re}$ number is based on equation
\ref{eq:Re_GL}.}
\end{table*}

Table~\ref{tab:measlocations} lists the physical measurement locations, where
the DNS values have been rescaled to $H=155~\textrm{mm}$.
Figure~\ref{fig:boundarylayerheights} shows that the wall distance for the
boundary layer measurements slightly deviates from linear behaviour (on a
double-log plot) due to minor changes in $\textrm{Pr}$ for the experimental points.
In separate experiments, LDA was used to determine mean-velocity and
rms-velocity profiles above the wall (at $\textrm{Ra} \approx 8.0\times 10^8$). The
results of these experiments confirmed that the measurement locations at
$z=z_\mathrm{bl}$ were all located in the boundary layer for the
full $\textrm{Ra}$ number range considered in this study.
The DNS values for $\lambda_u$, denoted by solid circles in
Fig.~\ref{fig:boundarylayerheights}, support this observation.
Clearly, (\ref{eq:blheight}) overestimates the scaling exponent of $\lambda_u$, so that the prefactor $\alpha$ in the BL measurement points $z_\textrm{\small bl} = \alpha \lambda_u$ is a decreasing function of $\textrm{Ra}$. Hence, as $\textrm{Ra}$ increases, the measurements are performed deeper into the kinematic boundary layer.
The disparity between the theoretical prediction (\ref{eq:blheight}) and the measurements for $\lambda_u$ will be discussed in more detail at the end of section \ref{par:spectra}.

The DNS time series are short in respect to the LDA series. To enable a
comparison between LDA and DNS results, the DNS data are extracted from four
points in the domain. These points are symmetrically placed around the origin
with horizontal coordinates $(\pm W/6,0)$, $(0,\pm W/6)$, as shown in
Fig.~\ref{fig:measpoints}. The four obtained time series are regarded as nearly
statistically independent, and are then averaged to effectively improve the
statistics of the DNS time series.

\section{Results}
\subsection{Probability density functions}
\begin{figure}
\subfigure[\label{subfig:varBL}boundary layer]{\includegraphics{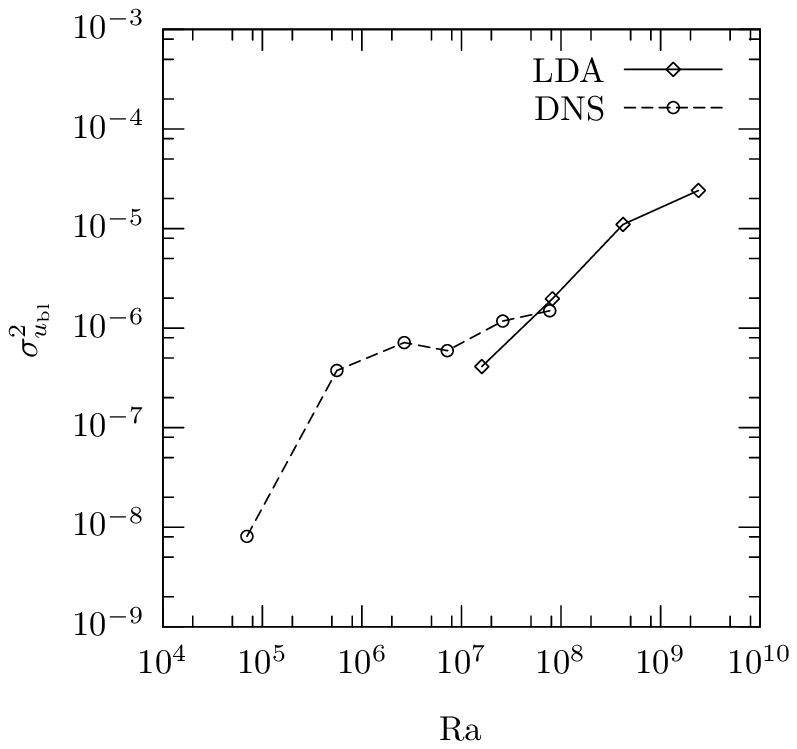}}
\subfigure[\label{subfig:varBulk}bulk]{\includegraphics{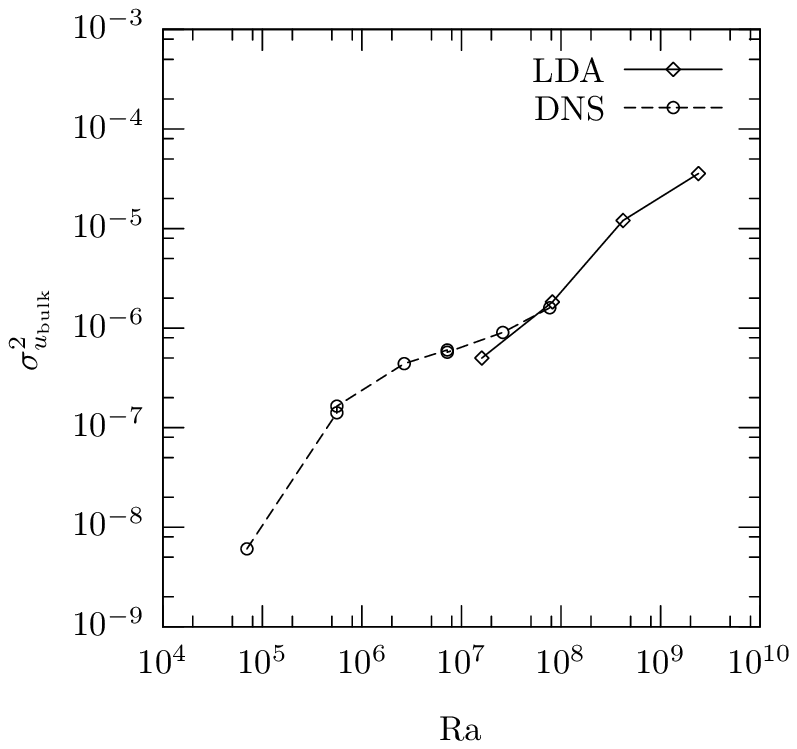}}
\caption{\label{fig:variances-bl}Variances of the horizontal velocity component,
$u$, in boundary layer~\subref{subfig:varBL} and in the
bulk~\subref{subfig:varBulk}.}
\end{figure}

Figure~\ref{fig:variances-bl} shows the measured and computed variance of the
horizontal velocity $\sigma^2_u$ in the boundary layer and in the bulk. The
graph shows that for increasing $\textrm{Ra}$ there is a gradual transition of the
variances between DNS and LDA results giving confidence in the procedure
followed.

\begin{figure}
\centering
\scalebox{0.7}{\subfigure[\label{subfig:pdfDNS_bl_T}boundary layer]{%
                \includegraphics{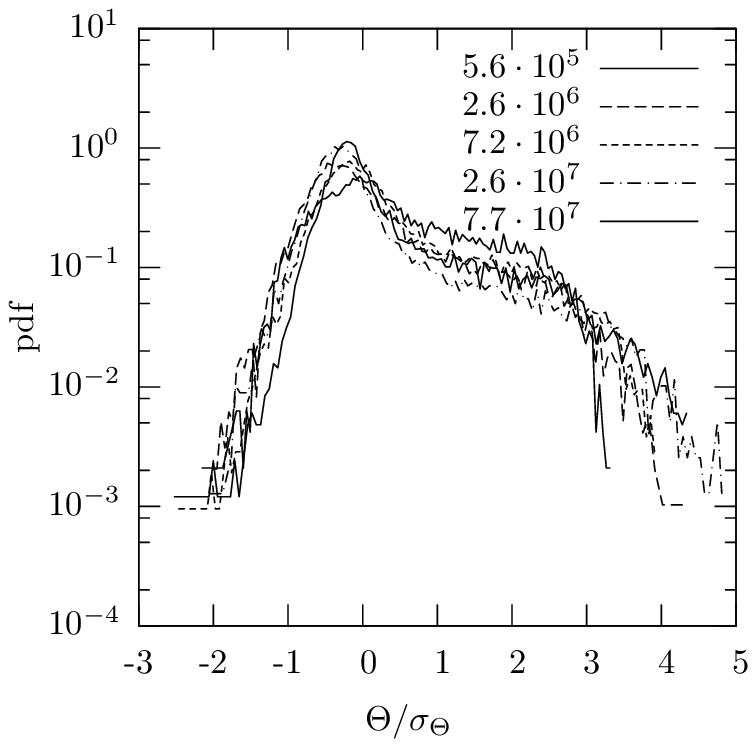}}}
\scalebox{0.7}{\subfigure[\label{subfig:pdfDNS_bulk_T}bulk]{%
                \includegraphics{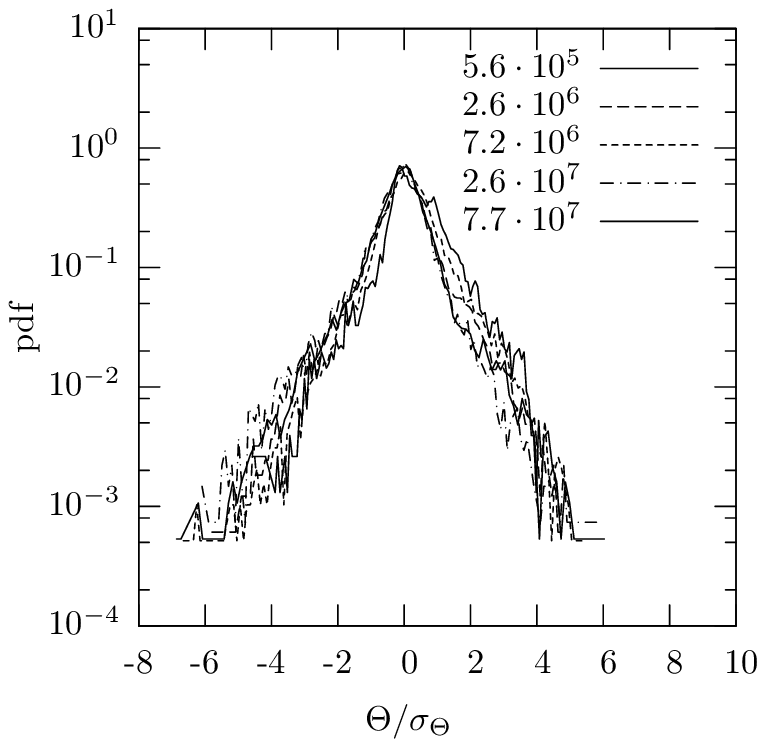}}}
\caption{\label{fig:pdfDNS_T}Probability density functions for
temperature $\Theta$ in the boundary layer \subref{subfig:pdfDNS_bl_T} and in
the bulk \subref{subfig:pdfDNS_bulk_T} of the RB cell obtained from
DNS. Histograms are scaled with their rms value $\sigma_\Theta$.}
\end{figure}

Temperature pdfs obtained from DNS for the boundary layer and the bulk are
depicted in Figure~\ref{subfig:pdfDNS_bulk_T} and show an exponential
distribution even for $\textrm{Ra}=5.6\times 10^5$. The straight lines (on logarithmic
scale) are generally associated with the hard turbulent regime, which sets in at
$\textrm{Ra}=4\times 10^7$ for aspect-ratio-1 cells (e.g.
\cite{Heslot1987,Castaing1989,Qiu2001}). Recall that the transition between soft
and hard turbulence takes place at relatively low $\textrm{Ra}$ number in the
aspect-ratio-4 cell, see \cite{Kerr1996}. The boundary layer pdf is skewed
towards positive temperatures. This is generally attributed to the interaction
of thermal plumes with the flow. Plumes leaving the boundary layer result in
positive temperature excursions.

\begin{figure*}
    \centering
    \begin{minipage}{\textwidth}
    \centering
    \subfigure[$\textrm{Ra}=7\times 10^4$ (DNS)]{%
        \label{subfig:pdfDNSRa7x104uv}
        \scalebox{0.7}{\includegraphics{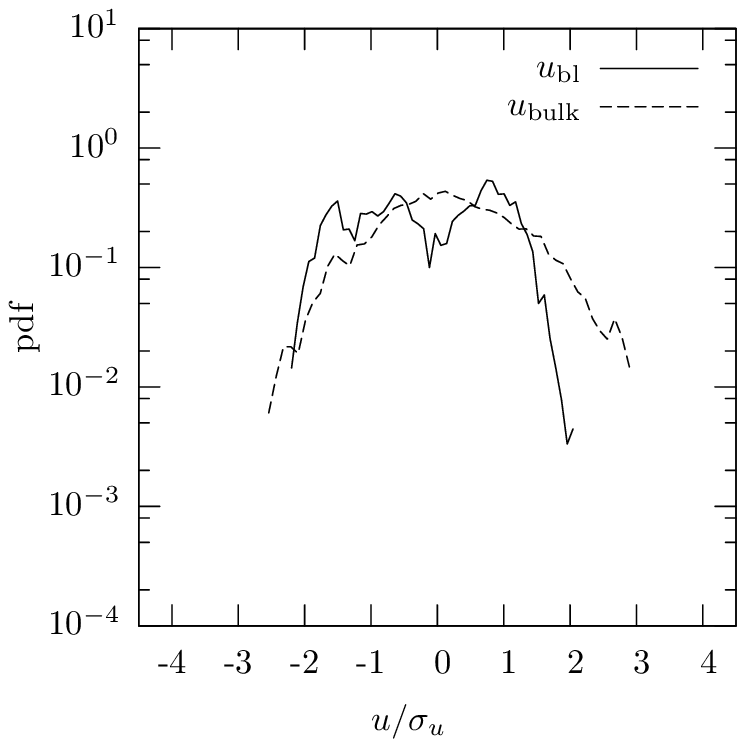}}
    }
    \subfigure[$\textrm{Ra}=5.6\times 10^5$ (DNS)]{%
        \label{subfig:pdfDNSRa6x105uv}
        \scalebox{0.7}{\includegraphics{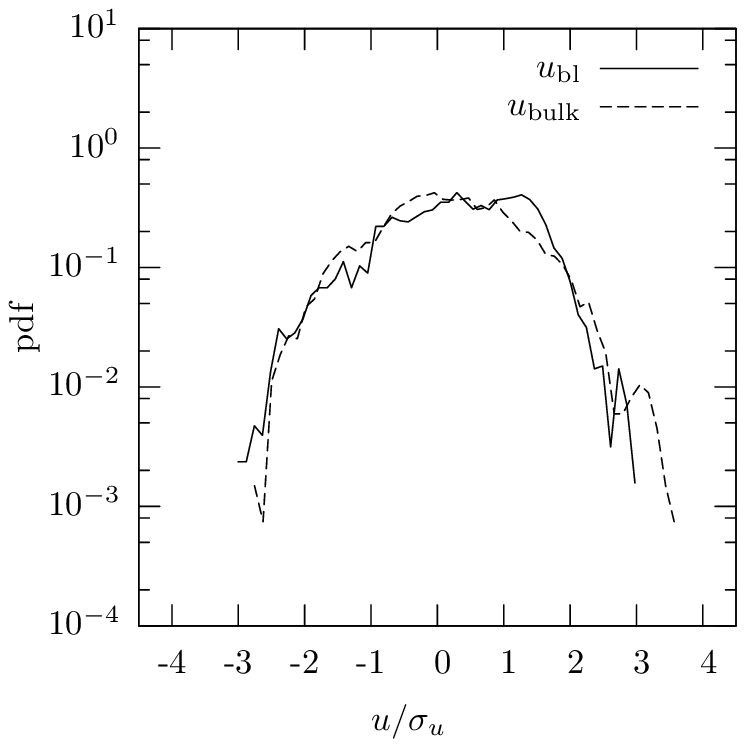}}
    }
    \end{minipage}
    \begin{minipage}{\textwidth}
    \centering
    \subfigure[$\textrm{Ra}=2.6\times 10^6$ (DNS)]{%
        \label{subfig:pdfDNSRa3x106uv}
        \scalebox{0.7}{\includegraphics{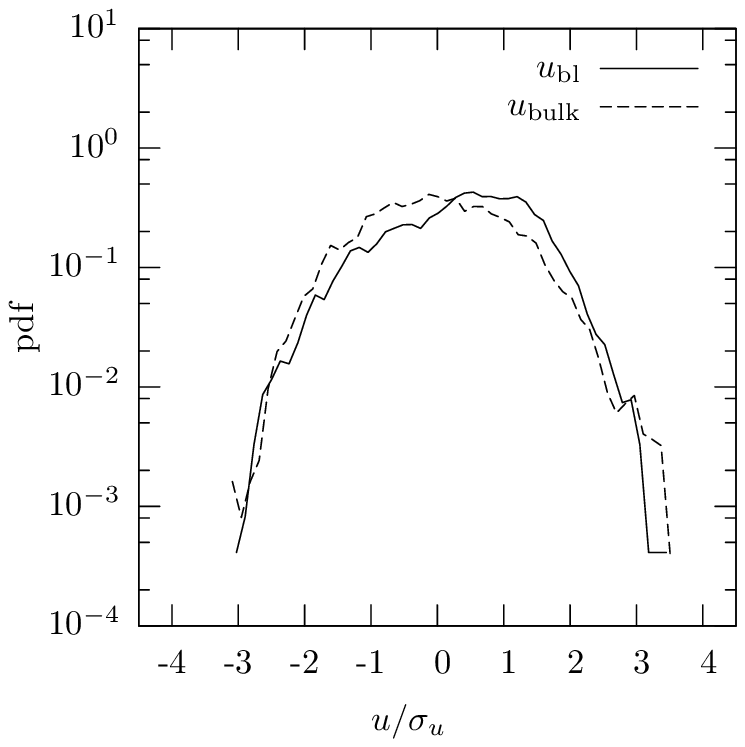}}
    }
    \subfigure[$\textrm{Ra}=7.2\times 10^6$ (DNS)]{%
        \label{subfig:pdfDNSRa7x106uv}
        \scalebox{0.7}{\includegraphics{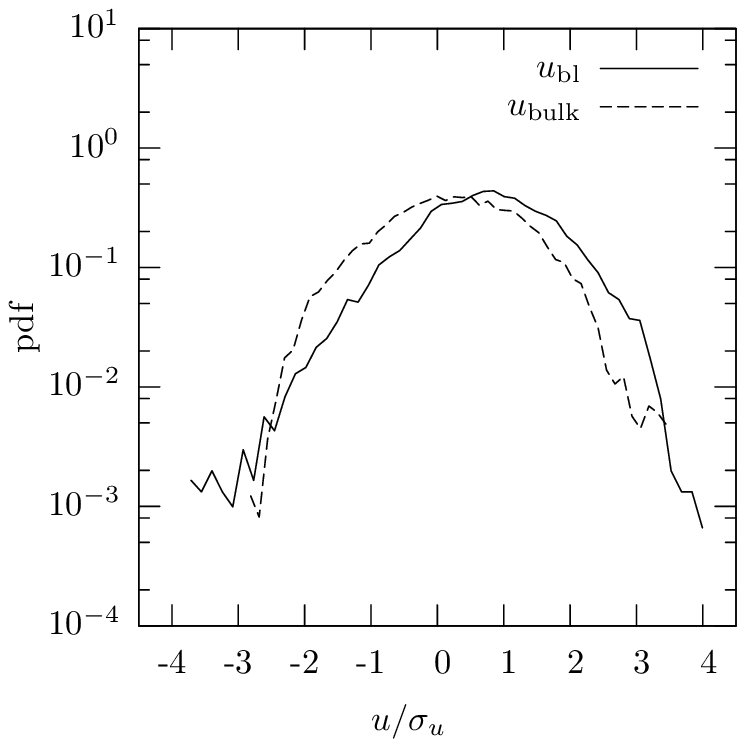}}
    }
    \end{minipage}
    \begin{minipage}{\textwidth}
    \centering
    \subfigure[$\textrm{Ra}=1.6\times 10^7$ (LDA)]{%
        \label{subfig:pdfLDARa2x107u}
        \scalebox{0.7}{\includegraphics{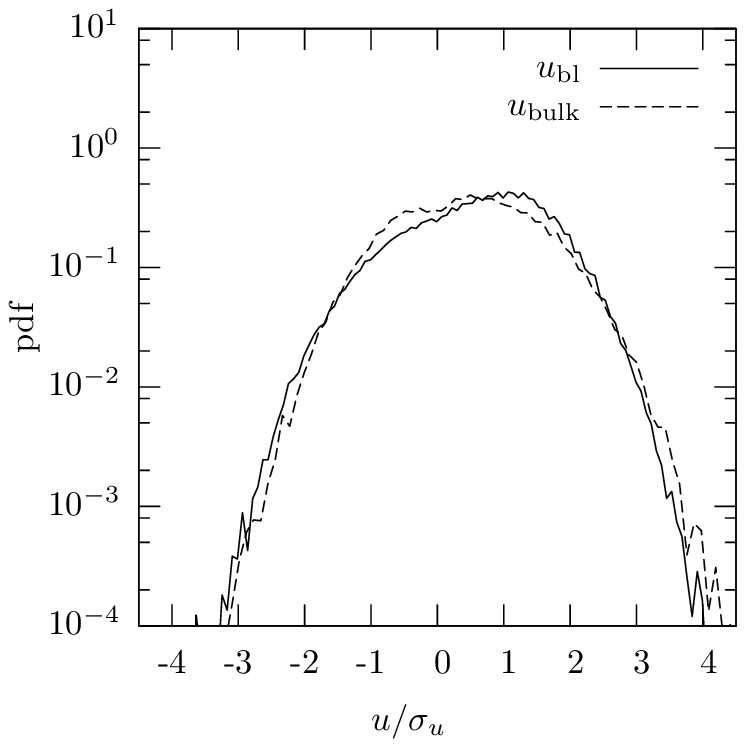}}
    }
    \subfigure[$\textrm{Ra}=2.6\times 10^7$ (DNS)]{%
        \label{subfig:pdfDNSRa3x107uv}
        \scalebox{0.7}{\includegraphics{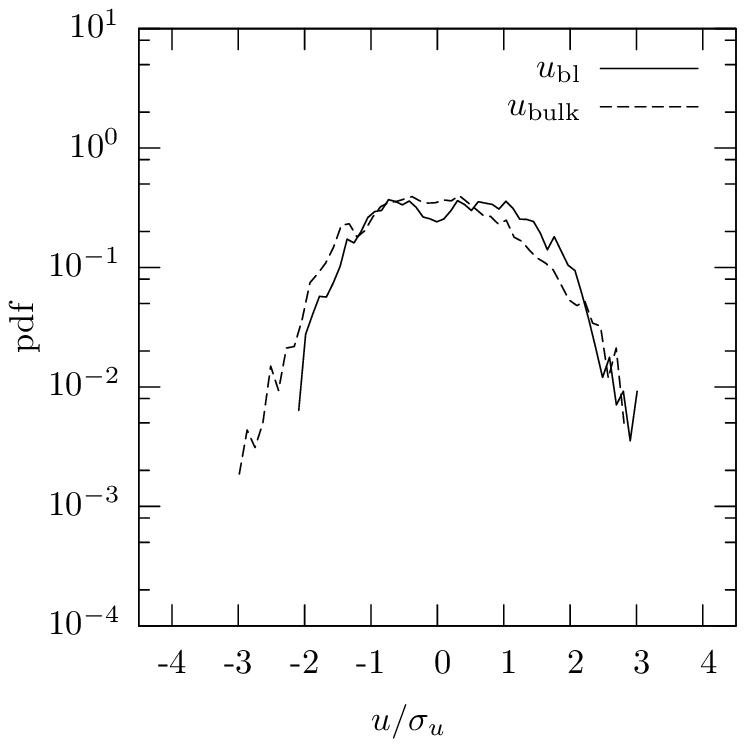}}
    }
    \end{minipage}
    \caption{\label{fig:pdfallblbulk_u}Probability density functions of the
$u$ horizontal velocity component in the boundary layer~(bl) and in the bulk of
the RB cell. Histograms are scaled with their rms value $\sigma_u$. The source
of the data, DNS or LDA, is indicated.}
\end{figure*}

Figure~\ref{fig:pdfallblbulk_u} depicts the probability density functions (pdfs)
of the computed and measured $u$ component. As expected, the pdfs of both
horizontal velocity components were found to have the same shape in the
simulations. For brevity only the $u$ component is therefore depicted in
Figure~\ref{fig:pdfallblbulk_u}. 
For a better comparison, the pdfs are scaled with the rms value $\sigma_u$. The
results show that the pdfs in both the boundary layer and in the bulk exhibit a
nearly Gaussian-like shape for the complete $\textrm{Ra}$ number range. Only the pdfs at
$\textrm{Ra}=7\times 10^4$ deviate largely from this trend, mainly due to the presence of
a strong oscillatory component in the velocity time series, as will be shown
later when discussing the power spectra.

\cite{Vincent1991,Mouri2002} make a distinction between sub-Gaussian, Gaussian
and hyper-Gaussian stages in the development of turbulence. They argue that the
nearly Gaussian shape of the pdfs in the present study indicates that turbulence
is in a fully developed stage. The pdfs for the highest $\textrm{Ra}$ values have
non-zero mean values. This is caused by the large-scale circulation or ``wind''
within the RB cell. For $\textrm{Ra}=4.2\times 10^8$ the mean value is approximately
$5~\mathrm{mm}\,\mathrm{s}^{-1}$. The wind is rather weak in the
sense that its mean value is small compared to the magnitude of the rms velocity
fluctuations, in accordance with other findings reported in
\cite{vanReeuwijk2005,Verdoold2006,Niemela2006,vanReeuwijk2007d}. It is also
observed that the wind is not always in the same direction. In some experiments
(Fig.~\ref{fig:pdfallblbulk_u}\subref{subfig:pdfLDARa8x107u}) it appears to be
clockwise, while in others 
(Figs.~\ref{fig:pdfallblbulk_u}\subref{subfig:pdfLDARa3x108u},
\subref{subfig:pdfLDARa2x109u}) it is counterclockwise.

\addtocounter{subfigure}{6}
\addtocounter{figure}{-1}

\begin{figure*}
    \begin{minipage}{\textwidth}
    \centering
    \subfigure[$\textrm{Ra}=7.7\times 10^7$ (DNS)]{%
        \label{subfig:pdfDNSRa7x107uv}
        \scalebox{0.7}{\includegraphics{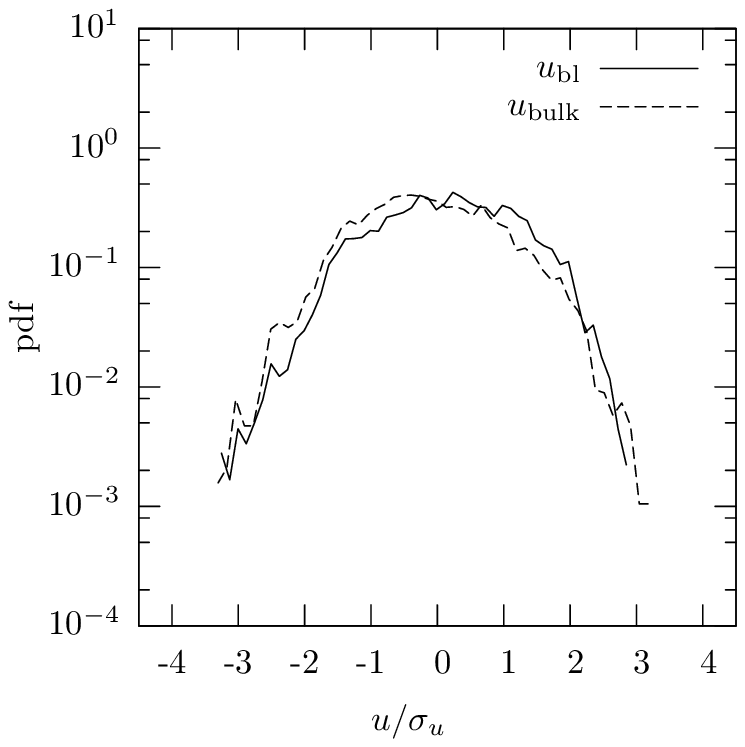}}
    }
    \subfigure[$\textrm{Ra}=8.2\times 10^7$ (LDA)]{%
        \label{subfig:pdfLDARa8x107u}
        \scalebox{0.7}{\includegraphics{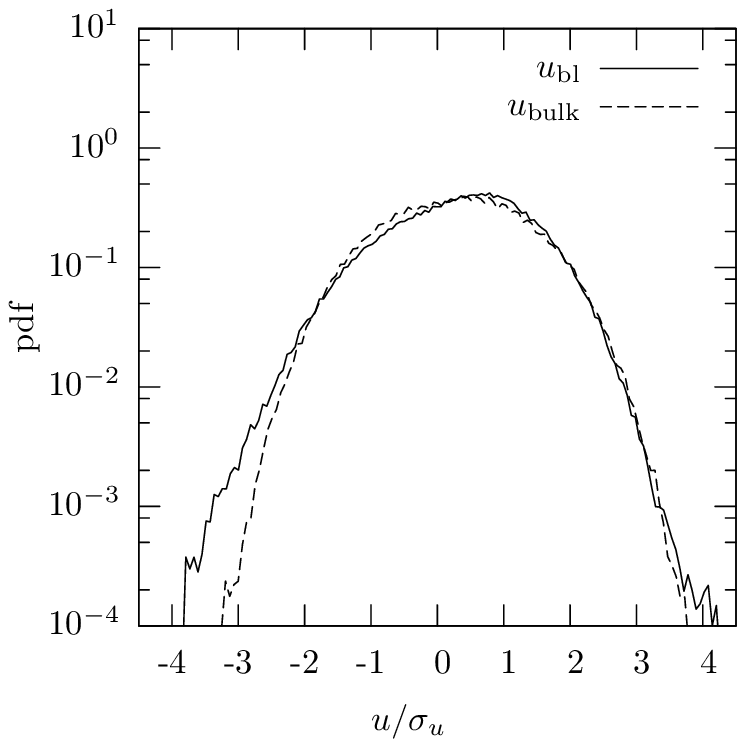}}
    }
    \clearpage
    \subfigure[$\textrm{Ra}=4.2\times 10^8$ (LDA)]{
        \label{subfig:pdfLDARa3x108u}
        \scalebox{0.7}{\includegraphics{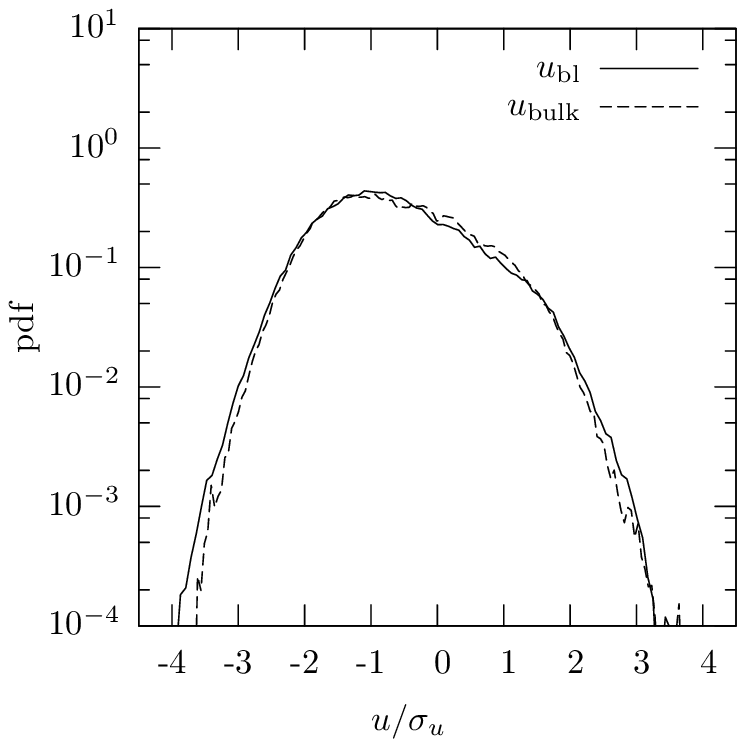}}
    }
    \subfigure[$\textrm{Ra}=2.4\times 10^9$ (LDA)]{
        \label{subfig:pdfLDARa2x109u}
        \scalebox{0.7}{\includegraphics{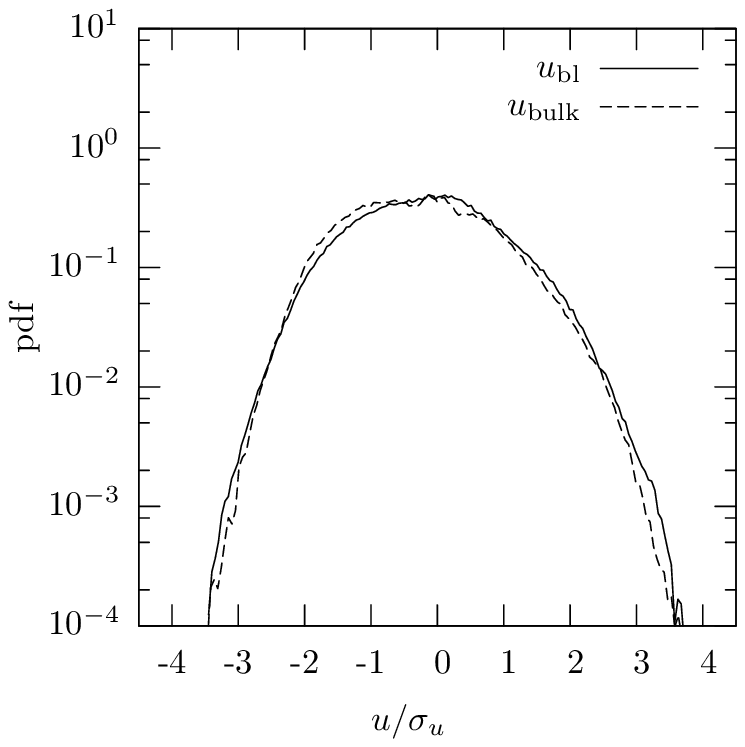}}
    }
    \end{minipage}
    \begin{center}
    Figure~\ref{fig:pdfallblbulk_u} (cont.)
    \end{center}
\end{figure*}

\addtocounter{figure}{1}
\setcounter{subfigure}{0}

\subsection{\label{par:spectra}Spectral density functions}
The obtained velocity time series are investigated further by computing the 
spectral density function (sdf). The sdf is defined as the Fourier transform of
the autocorrelation function $\rho(\tau)=\overline{u'(t) u'(t+\tau)} /
\overline{u'^2}$, where $u'(t)$ ($\equiv u(t)- \overline{u}$) is the fluctuating
component of the turbulent velocity, $t$ is time, $\tau$ is the time lag, and
the overline denotes ensemble averaging. All sdfs are calculated using the same
algorithm, to make a fair comparison for both DNS and LDA data. This algorithm,
described in \cite{Tummers2001}, can account for the randomness of the sampling
times in LDA measurements.

\begin{figure*}
    \centering
    \begin{minipage}{\textwidth}
    \centering
    \subfigure[$\textrm{Ra}=7\times 10^4$ (DNS)]{%
        \label{subfig:sdfDNSRa7x104}
        \scalebox{0.7}{\includegraphics{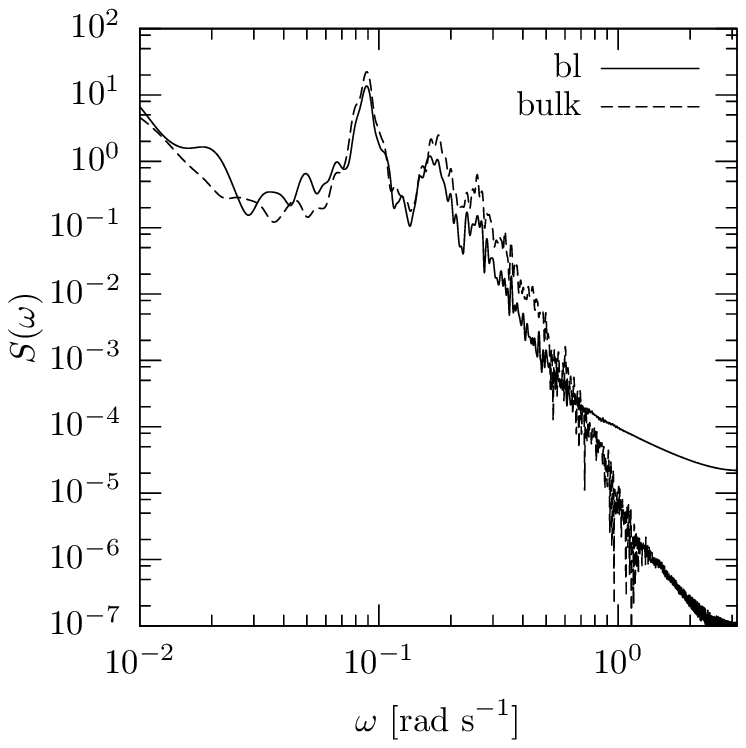}}
    }
    \subfigure[$\textrm{Ra}=5.6\times 10^5$ (DNS)]{%
        \label{subfig:sdfDNSRa6x105}
        \scalebox{0.7}{\includegraphics{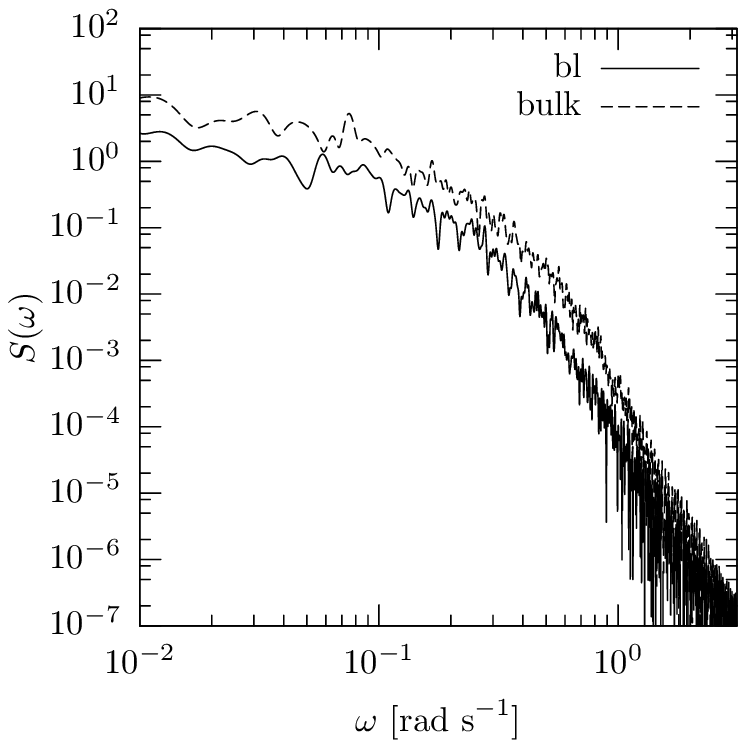}}
    }
    \end{minipage}
    \begin{minipage}{\textwidth}
    \centering
    \subfigure[$\textrm{Ra}=2.6\times 10^6$ (DNS)]{%
        \label{subfig:sdfDNSRa3x106}
        \scalebox{0.7}{\includegraphics{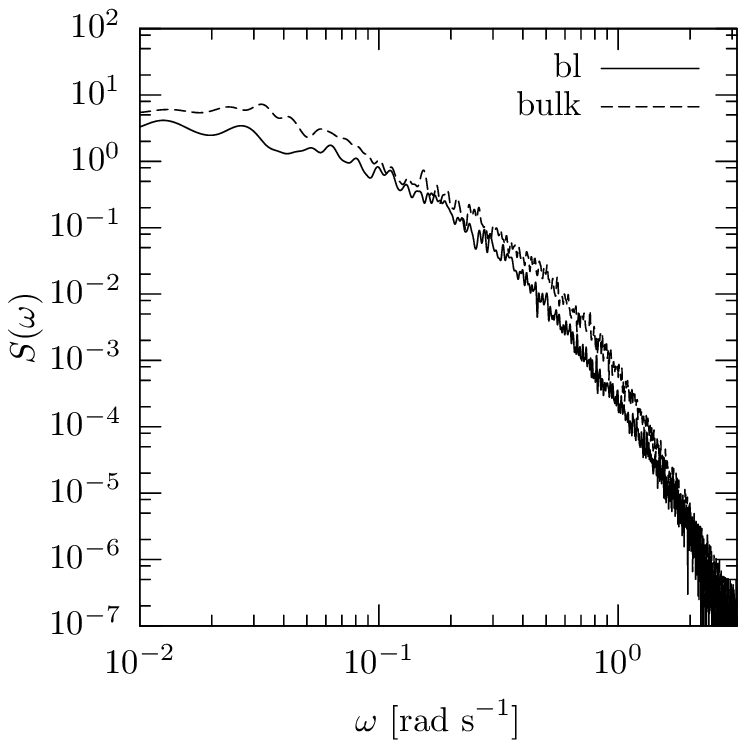}}
    }
    \subfigure[$\textrm{Ra}=7.2\times 10^6$ (DNS)]{%
        \label{subfig:sdfDNSRa7x106}
        \scalebox{0.7}{\includegraphics{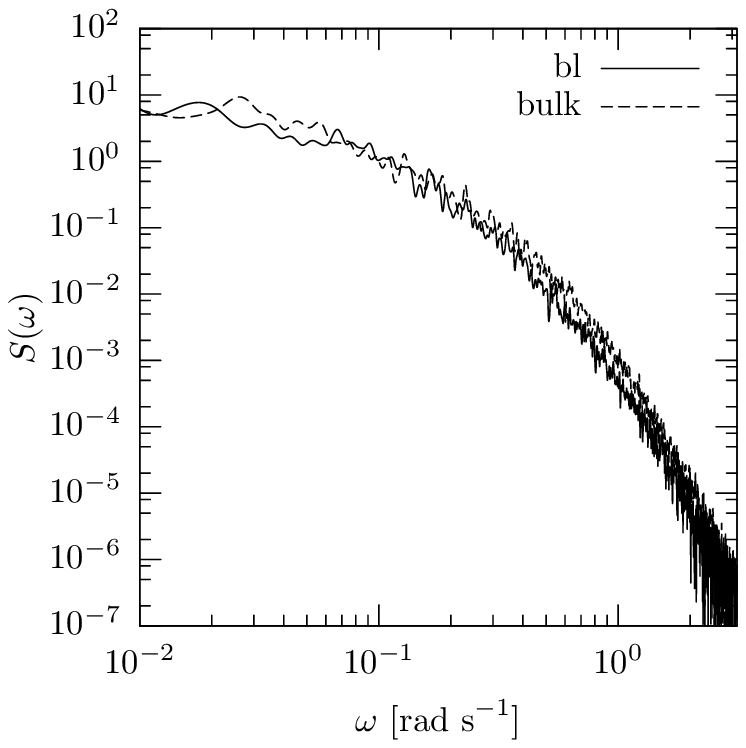}}
    }
    \end{minipage}
    \begin{minipage}{\textwidth}
    \centering
    \subfigure[$\textrm{Ra}=1.6\times 10^7$ (LDA)]{%
        \label{subfig:sdfLDARa2x107}
        \scalebox{0.7}{\includegraphics{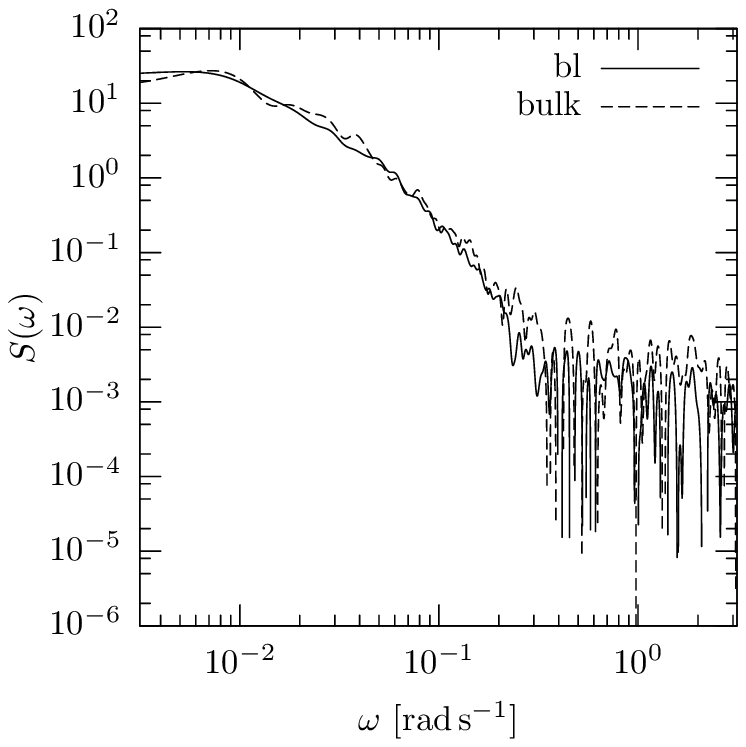}}
    }
    \subfigure[$\textrm{Ra}=2.6\times 10^7$ (DNS)]{%
        \label{subfig:sdfDNSRa2x107}
        \scalebox{0.7}{\includegraphics{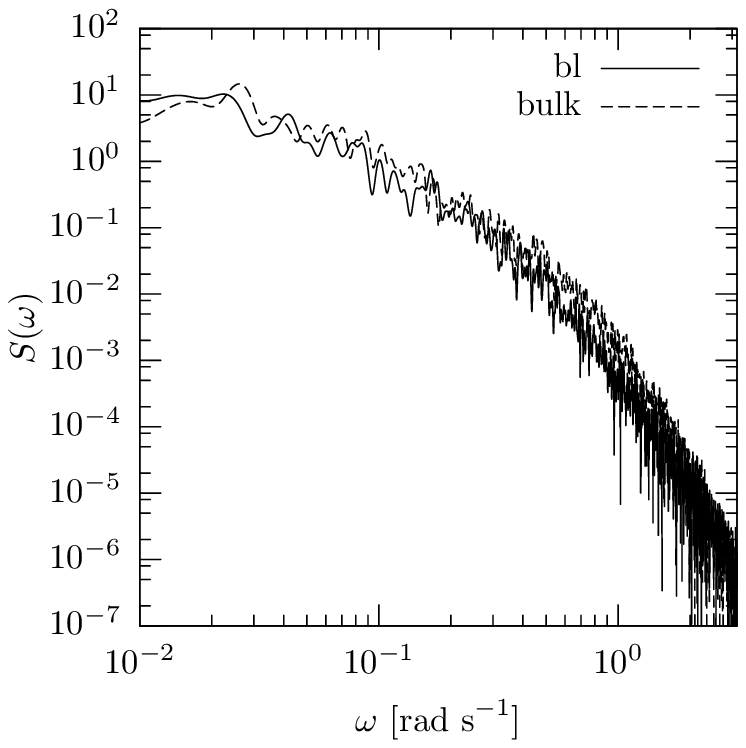}}
    }
    \end{minipage}
    \caption{\label{fig:spectrabulkbl}Spectral density functions for the
horizontal velocity component $u$ for points in the boundary layer~(bl) and in
the bulk of the RB cell. $\textrm{Ra}$ ranging from $7.0\times 10^4$ to $2.4\times 10^9$.
The source of the data, DNS or LDA, is indicated.}
\end{figure*}
\begin{figure*}
    \addtocounter{subfigure}{6}
    \addtocounter{figure}{-1}
    \begin{minipage}{\textwidth}
    \centering
    \subfigure[$\textrm{Ra}=7.7\times 10^7$ (DNS)]{%
        \label{subfig:sdfDNSRa8x107}
        \scalebox{0.7}{\includegraphics{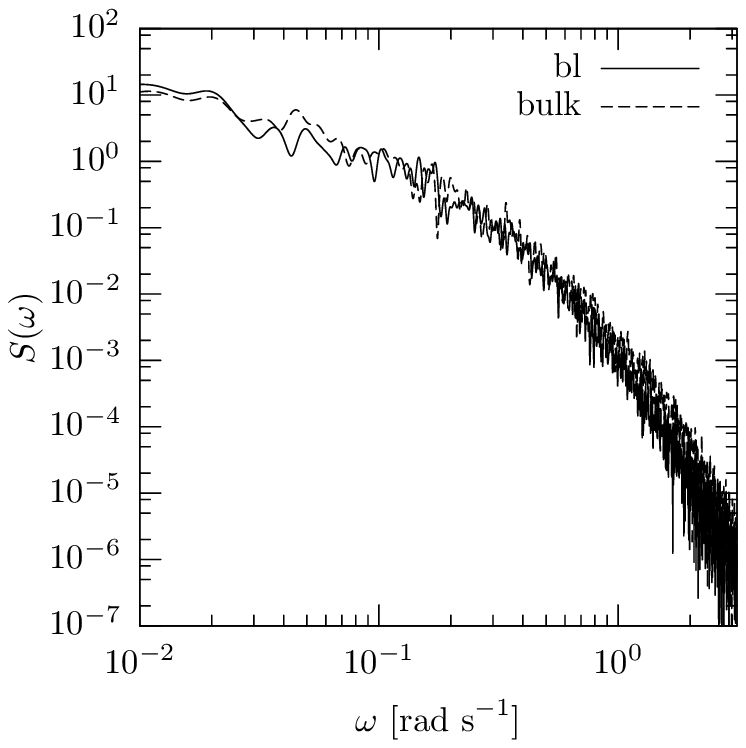}}
    }
    \subfigure[$\textrm{Ra}=8.2\times 10^7$ (LDA)]{%
        \label{subfig:sdfLDARa8x107}
        \scalebox{0.7}{\includegraphics{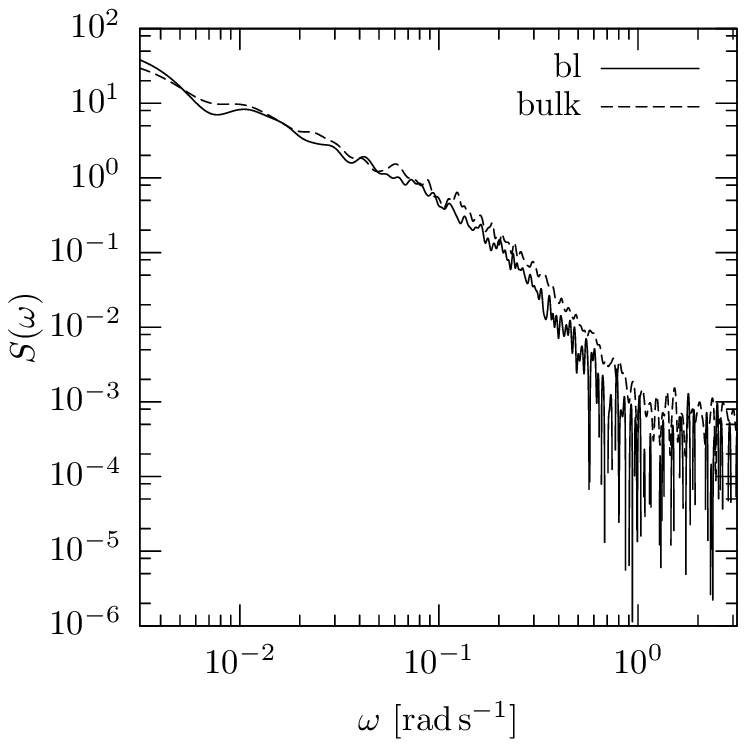}}
    }
    \end{minipage}
    \begin{minipage}{\textwidth}
    \centering
    \subfigure[$\textrm{Ra}=4.2\times 10^8$ (LDA)]{%
        \label{subfig:sdfLDARa3x108}
        \scalebox{0.7}{\includegraphics{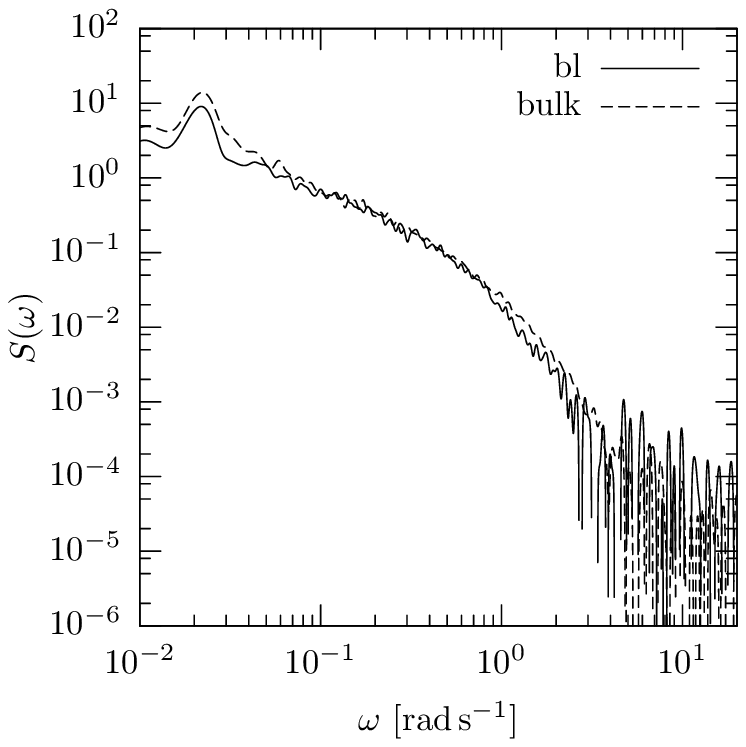}}
    }
    \subfigure[$\textrm{Ra}=2.4\times 10^9$ (LDA)]{%
        \label{subfig:sdfLDARa2x109}
        \scalebox{0.7}{\includegraphics{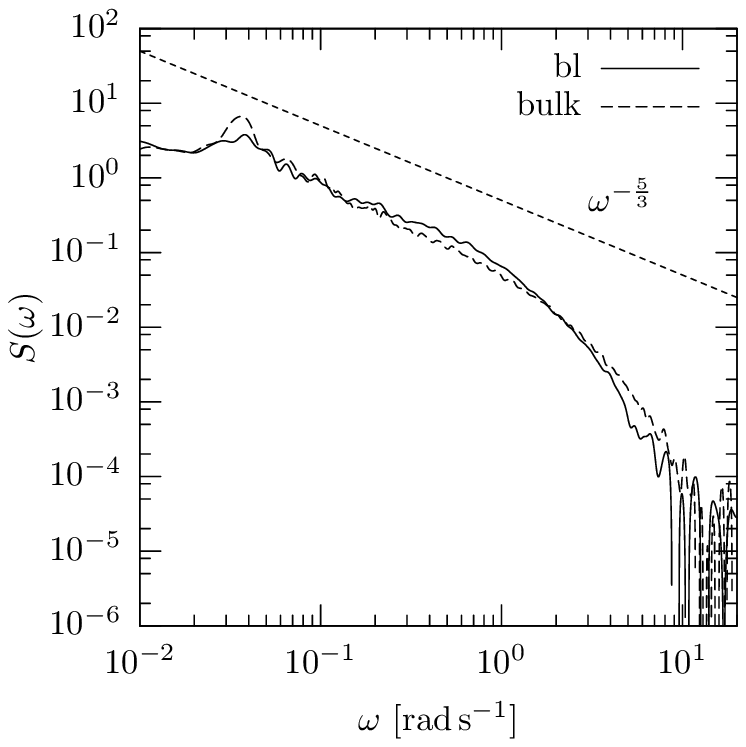}}
    }
    \end{minipage}
    \begin{center}
    Figure~\ref{fig:spectrabulkbl} (cont.)
    \end{center}
\end{figure*}

Figure~\ref{fig:spectrabulkbl} shows the sdfs of the horizontal component in the
$x$-direction determined in the boundary layer and in the bulk for the full
range of $\textrm{Ra}$ considered in this study. The sdfs determined from DNS data are
accurate at high frequencies, but less reliable in the low frequency range due
to the relatively small measurement times. In contrast, the low frequency region
in the LDA sdfs is accurate, while the high frequency range suffers from a
relatively high statistical scatter caused by the random sampling process in
LDA. The level of the statistical scatter (which strongly depends on the mean
data rate) sets a lower limit to the spectral density that can be accurately
determined from LDA data.

When comparing the various sdfs in Figure~\ref{fig:spectrabulkbl} it is clear
that the sdf for $\textrm{Ra}=7\times 10^4$ is different from all others. At this low
$\textrm{Ra}$ numbers, the flow is not turbulent but in a state of spatio-temporal chaos.
As mentioned earlier, the velocity time series is characterised by strong
oscillations. As a consequence, the sdf in Fig.~\ref{subfig:sdfDNSRa7x104} (and
to a lesser extent the sdf in Fig.~\ref{subfig:sdfDNSRa6x105}) is dominated by a
limited number of discrete peaks and the sdf rolls off rapidly at higher
frequencies.

Peaks in the sdf also occur in the very low frequency range in
Figs.~\ref{subfig:sdfLDARa3x108} and \ref{subfig:sdfLDARa2x109}. The peak
frequencies $\omega_\mathrm{p}$ correspond to periods of $2 \pi/
\omega_{\mathrm{p}}=283$~s and $172$~s at $\textrm{Ra}=4.2 \times 10^8$ and $\textrm{Ra}=2.4
\times 10^9$, respectively. These oscillations are unrelated to the ones
observed at low $\textrm{Ra}$, and have their origin in the alternating growth and decay
of rolls that form the large scale circulation as discussed in
\cite{Verdoold2006}.

For $\textrm{Ra}> 2.6 \times 10^6$ all sdfs measured in the bulk have a wide range of
active scales, and this range increases with $\textrm{Ra}$. These sdfs are continuous,
and are not dominated by a small number of frequencies as one might expect in
spatio-temporal chaos. Clearly these sdfs indicate that the bulk in the RB cell
is turbulent for $\textrm{Ra}> 2.6 \times 10^6$. Although an inertial subrange is absent
for the measurements at lower $\textrm{Ra}$, the sdf at $\textrm{Ra}=2.4 \times 10^9$ (Fig.\
\ref{subfig:sdfLDARa2x109}) shows an inertial subrange spanning roughly one
decade. Note that the presence of an inertial subrange is not a necessary
condition for turbulence. An inertial subrange can only form when there is a
range of scales which is neither influenced by macro- or micro-scales, i.e.\ at
high $\textrm{Re}$ when the production and dissipation scales are sufficiently separated.
Therefore, an inertial subrange is absent despite the turbulence at lower $\textrm{Re}$
\cite[see also][]{Wu1990, Camussi1998}.

The most striking observation is that the sdfs in the bulk and in the boundary
layer are practically identical for $\textrm{Ra}> 2.6 \times 10^6$. As the sdfs for the
bulk indicate that the flow is turbulent, one cannot escape from the conclusion
that the flow in the boundary layers must also be characterised as turbulent. It
is difficult to reconcile this observation with the conceptual image of a
laminar, Blasius-type boundary layer along the walls of the RB cell. For a
Blasius boundary layer, the laminar-turbulent transition occurs at a relatively
high Reynolds number, i.e., $\textrm{Re}^*_c \approx 10^5-10^6$ depending on the
intensity and nature of disturbances in the flow \citep{Schlichting1997}. Here,
$\textrm{Re}^*_c$ is a Reynolds number based on plate length and outer velocity. However,
even at the highest $\textrm{Ra}$ considered in this study, the value of $\textrm{Re}^*$ based on
the cell width and the wind velocity is only $3 \times 10^3$, which is far below
the critical value for laminar-turbulent transition in a flat plate boundary
layer. Thus, the turbulence in the boundary layer cannot be due to a
shear-triggered transition (occuring inside the boundary layer), but more likely
due to the impingement and detachment of plumes (travelling into and out of the
boundary layer), which is corroborated by the great similarity between the bulk
and the boundary-layer spectra.

Naturally, one can always envision that the boundary layer only
\emph{quantitatively} behaves as a laminar Blasius boundary layer: the plumes
introduce a time-dependence on time-scales that prevent pure laminarity. The
underlying assumption is that on the average, the thermals and plumes are
passive when it comes to scaling of various important integral parameters of the
kinematic boundary layer such as the friction factor $C_f$ and the boundary
layer thickness $\lambda_u$. Indeed, $C_f$ scales as $C_f \propto \textrm{Re}^{-1/2}$, as
one would expect for a laminar boundary layer past a flat plate at moderate $\textrm{Ra}$
\citep{Chavanne1997, Chavanne2001, Amati2005, vanReeuwijk2007c}. 

The scaling of the boundary layer thickness $\lambda_u$ is slightly
more complicated. Assuming a laminar boundary layer, $\lambda_u$ scales as
$\lambda_u \propto \textrm{Re}^{-1/2}$ \citep{Schlichting1997}, which is related to $\textrm{Ra}$
as $\lambda_u \propto \textrm{Ra}^{-1/4}$ upon assuming $\textrm{Re} \propto \textrm{Ra}^{1/2}$.
Experiments show that a scaling exponent $-0.25$ holds well for the boundary
layer thickness of the sidewalls \citep{Qiu1998}, but not for the scaling of the
bottom and top boundary layers \citep{Qiu1998a, Kerr1996, vanReeuwijk2007c}
which scales with exponents from $-0.11$ to $-0.16$. 
The simulations in this paper confirm the weak $\textrm{Ra}$ number dependence of
$\lambda_u$ (Fig.\ \ref{fig:boundarylayerheights}).
This disparity indicates that plumes impinging
on and emerging from the boundary layer are not entirely passive, but actively
influence the boundary layer thickness. Given the important role of $\lambda_u$
in the theory of \cite{Grossmann2000}, it seems to us that a detailed study of
how the interplay of plumes, the wind and the boundary layer influences
$\lambda_u$ is desirable. A first step in this direction is taken in a
forthcoming study \cite{vanReeuwijk2007d, vanReeuwijk2007c}, of which the
results indicate that the scaling of $\lambda_u$ cannot be derived from laminar
boundary layer theory and scales as $\lambda_u \propto \lambda_\Theta^{-1}
\textrm{Re}^{-1}$.

Although the sdfs point towards the presence of turbulence in the boundary
layer, the characteristics of the boundary layer are of a rather different
nature than those of classical (i.e.\ shear-driven) turbulent boundary layers.
Indeed, viscous forces are still significant in the boundary layer, as can be
deduced from the domain height $H$ in viscous units $\delta_v = \nu / u_\tau$,
where $u_\tau \equiv \sqrt{\nu \frac{\partial u}{\partial z} |_\mathrm{w}}$ is
the friction velocity \cite{Schlichting1997}. Assuming that
$\frac{\partial u}{\partial z}|_\mathrm{w} \approx U / \lambda_u$, the
ratio $H / \delta_v$ which is the shear-Reynolds number $\textrm{Re}_\tau$, can be
approximated by
\begin{equation}
  \textrm{Re}_\tau \equiv \frac{u_\tau H}{\nu} \approx
       \left(\frac{\lambda_u}{H}\right)^{-1/2} \textrm{Re}^{1/2}.
\end{equation}
The simulations show that $\textrm{Re}_\tau = 70$ at $\textrm{Ra}=2.1 \times 10^7$, and the
viscous thickness of the boundary layer $\lambda_u/\delta_v = \lambda_u / H
\textrm{Re}_\tau$ is only a fraction of this already very low value. Therefore, viscous
effects are significant in the boundary layer.

It is interesting to note that the scaling of the boundary layer thickness for
flat plate flow changes from $\delta / x \propto \textrm{Re}_x^{-1/2}$ (laminar) to
$\delta / x \propto \textrm{Re}_x^{-1/7}$ (turbulent, see \cite{White1991}), where $\textrm{Re}_x$
is the Reynolds number based on the distance $x$ to the leading edge of the
plate, and the thickness $\delta$ is the location where the velocity is 99\% of
the outer velocity. 
The decreased exponent resembles larger entrainment and thus a quicker growth of
the boundary layer as a function of $x$. 
Upon assuming that the typical development length is $H$ and that $\textrm{Re} \propto
\textrm{Ra}^{1/2}$, the anticipated scaling for a turbulent boundary layer is $\delta / H
\propto \textrm{Ra}^{-1/14}$. 
If $\alpha \approx -0.11$ is the scaling exponent $\lambda_u \propto \textrm{Ra}^\alpha$
from the simulations, then $-1/4 < \alpha < -1/14$, so the exponent may resemble
an intermediate state.

However, it is not clear whether the entrainment mechanism of a developing
boundary layer past a flat plate is transferable to RB convection; the active
role of plumes which transfer fluid (and heat) into and out of the boundary
layer may significantly affect the entrainment
characteristics. It is our suspicion that the relatively weak forcing in the
horizontal direction due to the wind (which is responsible for the friction
factor $C_f$) is only weakly coupled to the relatively strong forcing in the
wall-normal direction due to buoyancy. This may be the reason why the boundary
layer has an sdf which is typical of developed turbulence but has many laminar
features at the same time. Due to the weak coupling, the conceptual image of a
laminar boundary layer may be appropriate for $C_f$ but not for $\lambda_u$ in
the boundary layer in RB convection. Therefore, the straightforward application
of well-known characteristics of the Blasius-type boundary layer (such as the
criteria for laminar-turbulent transition and the scaling of $\lambda_u$ with
$\textrm{Re}$) to the boundary layers in RB convection is questionable without additional
justification.

\section{Conclusion}
A combined numerical and experimental investigation was carried out to study the
boundary layers in RB convection in a 4:1 aspect-ratio cell filled with water.
The study covers a wide range of $\textrm{Ra}$ numbers varying between $7.0 \times 10^4$
and $2.4 \times 10^9$. The results for the lower $\textrm{Ra}$ numbers were obtained from
direct numerical simulations, while those for the higher $\textrm{Ra}$ numbers followed
from laser Doppler anemometry. In the midrange both techniques have an overlap
of about one decade.

The probability density functions of the horizontal velocity components have a
Gaussian-like shape for the complete $\textrm{Ra}$ number range, except at the lowest
$\textrm{Ra}$ where the flow is in a state of spatio-temporal chaos. Probability density
functions that are scaled with the velocity rms value overlap for the entire
$\textrm{Ra}$ number range, with the exception of the largest $\textrm{Ra}$ numbers where the
large-scale circulation, or ``wind'', is predominantly visible in the
histograms.

The spectral density functions of the horizontal velocity component that were
determined in the bulk flow and in the boundary layer are surprisingly similar.
Spectra in both regions of the flow are continuous and have a wide range of
active scales for $\textrm{Ra}>2.6 \times 10^6$. This indicates that both the bulk and
the boundary layers are turbulent in the $\textrm{Ra}$ number range considered, while (i)
molecular effects are still noticeable in the boundary layer and (ii) the
boundary layer does not behave like a classical shear-driven turbulent boundary
layer.

\bibliography{paper}
\bibliographystyle{apsrev}
\end{document}